# Force spectroscopy in studying infection

Zhaokun Zhou, Mark C. Leake

Abstract

Biophysical force spectroscopy tools – for example optical tweezers, magnetic tweezers, atomic force microscopy, – have been used to study elastic, mechanical, conformational and dynamic properties of single biological specimens from single proteins to whole cells to reveal information not accessible by ensemble average methods such as X-ray crystallography, mass spectroscopy, gel electrophoresis and so on. Here we review the application of these tools on a range of infection-related questions from antibody-inhibited protein processivity to virus-cell adhesion. In each case we focus on how the instrumental design tailored to the biological system in question translates into the functionality suitable for that particular study. The unique insights that force spectroscopy has gained to complement knowledge learned through population averaging techniques in interrogating biomolecular details prove to be instrumental in therapeutic innovations such as those in structure-based drug design.

1. Introduction

Force spectroscopy refers to a series of experimental techniques capable of linear and rotational force (the latter also referred to as torque) transduction onto and the measurement of microscopic length scale objects and smaller, biological samples in particular. The forces can arise from the transfer of photon momenta (optical tweezers), magnetic interaction between a generated *B* field and a magnetic micro-object (magnetic tweezers), mechanical manipulation (atomic force microscopy, micropipette and microneedle) and liquid flow pressure. The samples are usually imaged or tracked with optical microscopy – brightfield, fluorescence imaging and so on. But the configuration of force spectroscopy instruments have uniquely allowed other measurement methods, such as back focal plane interferometry. These methods are explained below. With its capability to transduce force to perturb the conformation of the biological samples, force spectroscopy has found applications in studying elasto-mechanical properties of biological polymers such as the over-stretching (Smith et al. 1996), supercoiling (Strick et al. 1996), braiding (Strick et al. 1998), buckling (Strick et al. 1996) and denaturing (King et al. 2013) of nucleic acids and torsional compliance of bacteria flagella (Block et al. 1989), and probing muscle proteins for shock-absorbing properties (Leake et al. 2003; Leake et al. 2004; Linke and Leake 2004; Leake et al. 2006; Bullard et al. 2006) (REFS), all techniques which uses physics to probe biology one molecule at a time (Leake 2013). Another advantage afforded by these techniques stems from the fact that they study one copy of the sample at a time, be it a biological macromolecule, a cell or a virus, which, in contrast to ensemble methods that have studied biological molecules and organisms prior to the advent of force spectroscopy, does not need to temporally and spatially average over population measurements. As a result, force spectroscopy techniques can interrogate the time-resolved series of molecular actions that comprise a complete catalytic or translocation cycle. Among the achievements are the characterisation of topoisomerase actions in uncoiling double stranded nuclei acids (Charvin et al. 2005) and the walking patterns of myosin (Finer et al. 1994) on actin fibres and kinesin (Block et al. 1990)/dynein (Gennerich et al. 2007) on cytoskeletal microtubules. Also worth mentioning is the increasing spatial-temporal resolution achieved over the years since the inception of force spectroscopy. Super-resolution fluorescence microscopy and interferometry position tracking with split photodiode detection afford sensitive detection whilst continual improvements in the instrumental design – acoustic and mechanical vibration isolation, atmospheric air purification by replacement with helium gas, active feedback sample drift cancelling and so on – minimize instrumental noise to fundamental sources such as Brownian noise and shot noise. Angstrom resolution has already been achieved, which is necessary to resolve movement on, for example, DNA double strands with base pair repeating distances of 3.4 A. This opens the door to the observation of nucleic acid motor proteins: topoisomerases, helicases (Fili et al. 2010), gyrases (Gore et al. 2006) and RNA polymerases (Abbondanzieri et al. 2005), etc.

Questions linked to infection have been studied with force spectroscopy experiments to advance our understanding in infection so we can more effectively prevent, diagnose and treat infectious diseases. These experiments have mostly focused on molecular and cellular levels but they involve many aspects of infection including the metabolism and behaviour of infectious microorganisms, the effect of drugs on infectious microorganisms, the properties of our immune cells and so on.

Heterogeneity among individual cells can arise as a response to uneven substrate distributions, as a result of detecting different signal molecules, or simply from stochastic variations. Population averaging measurements are unable to distinguish between statistical distributions, such as between normal and bimodal distributions, both common among cell expressions. As Lidstrom *et al*. pointed out, normally distributed phenotypes imply a cooperative binding mechanism while a bimodal distribution imply an on/off switch mechanism determined by a threshold level with some stochastic elements (Lidstrom and Konopka 2010). In contrast, information pertaining to the state of every cell allows the determination of the distribution and thus the biological mechanism. Similar necessity for single entity measurements is seen in cell responses, as indicated by the level of some measurable parameters, to perturbations. Immediate response from part of the population to a perturbation has the same ensemble appearance as a slow or delayed response from the entire population, which again calls for the measurement of single individuals.

Structure-based drug design is a directed and systematic approach of discovering new drugs that does not resort to traditional trial-and-error testing of myriads of candidate chemicals. Rather, the structural features that enable the drug to block, inhibit or activate a target protein are designed according to the structural features on the target protein. So it is crucial to possess detailed and accurate knowledge of the target proteins. Traditional ensemble methods to investigate structural properties of proteins - biochemical studies, NMR, mass spectroscopy, X-ray crystallography - as well as the traditional single-molecule imaging method electron microscopy, have had great successes in discovering molecular structures which has led to breakthroughs in drug development. However, during a catalytic cycle or force generation action, the conformation of or charge distribution on a protein may change throughout the process, taking a few distinctive forms. The aforementioned techniques are limited by the necessity of sample preparations that isolate and fix proteins so the rich conformational states will not be sufficiently present during detection. In contrast, force spectroscopy tools allow proteins to complete a set of actions as they do in living, physiological settings so the change of states at each phase of the process can be characterised with the high spatial-temporal resolution afforded by force spectroscopy tools. Apart from the ability to characterise molecular structures during molecular activity, force spectroscopy can also characterise the molecules' hindered actions when they are inhibited either by naturally existing chemicals or manmade drugs. Naturally existing inhibitory chemicals may be secreted by a nearby organism to suppress competitors. The molecular details of the inhibitory mechanism will shed light on the response of molecules to the binding of the inhibitor and inspire the design of new drugs. Manmade drugs need to be assessed for the effectiveness and the current main approach is *in silico* simulation, which is sometimes prohibited by heavy computational loads. Direct empirical observations will fill in the gaps where simulations are infeasible.

This chapter will summarise the designs and measurement mechanisms of three examples of force spectroscopy instruments and their applications in studying biological agents which, among other achievements, tremendously advance our ability at understanding infectious diseases.

2. Introduction to force spectroscopy tools

Optical tweezers, magnetic tweezers and atomic force microscopy, among others, are three types of microscopic force transduction instruments capable of applying controlled forces and torques onto single molecules and cells as a means of perturbing the biological samples to reveal their mechanical and/or dynamical properties. Some of them also serve as a contact-mode imaging technique with molecular precision. Their mechanisms and recent advances have been described in detail in literature (Neuman and Nagy 2008; Moffitt et al. 2008; De Vlaminck 2012).

2.1. Optical tweezers (OT)

Photons carry momenta and when they enter a medium of different refractive index, they deflect and impart part of their momenta to the new medium – this serves as the underlying principle of OT in force application to microscopic length scale objects. A collimated laser beam is typically focused via a microscope objective lens to form a diffraction limited spot so that light rays converge strongly in all three dimensions towards the centre of the spot (fig. 1 a, b and e). Any microscopic object of higher refractive index than the surrounding medium will experience gradient forces that push it to the centre of the trap (fig. 1 c and d) – a ray trace diagram can show this (Ashkin 1992). Since the gradient force scales linearly with light intensity, the maximum force is limited by the sample heating at high laser powers. But this is greater than many biological forces at molecular and cellular levels. OT routinely achieves sub-nanometre and sub-millisecond resolutions since the optical system of OT is compatible with a quadrant photodiode detection that images the interference pattern due to the bead in the beam waist formed at the back focal plane of the condenser lens.

2.2. Magnetic tweezers (MT)

A magnetic dipole $\boldsymbol{m}$ in a magnetic field $\boldsymbol{B}$ experiences a magnetic force $\boldsymbol{f}$ (fig. 2c) and torque $\boldsymbol{\tau}$ (fig. 2d): the force acting on the dipole is proportional to the field gradient: $\boldsymbol{f} = (\boldsymbol{m} \cdot \nabla)\boldsymbol{B}$ (Shevkoplyas et al. 2007) and points along the direction of the gradient; the torque scales with the field itself: $\boldsymbol{\tau} = \boldsymbol{B} \times \boldsymbol{m}$ (Mosconi 2011) and points in the direction of the field. MT uses this principle to trap and rotate magnetic particles: a $\boldsymbol{B}$ field is generated either with permanent magnets (fig. 2a) or electromagnets (fig. 2b) and a microscopic magnetic object (called the 'handle') is placed inside the field. Thus all translational and rotational degrees of freedom of the handle can be controlled. The biological sample is usually covalently attached to the handle and any translation and rotation of the bead are passed on to the sample. The magnetic force can easily reach biologically relevant values of tens of pN: titin unfolds at applied force of 20 to 30 pN (Kellermayer et al. 1997) and double-stranded DNA overstretches at 65 pN (Smith et al. 1996). The minimum of applicable force is just as important since it determines whether the smallest biological force can be measured. In the case of MT, this value can be as low as thermal forces. The ease with which MT applies torque and twists is what lends MT its popularity. Compared to optical tweezers, the creation of a stable and low-noise rotating magnetic field is economical and technically straightforward. Biological values of torque are in the range of a few to a few tens of pN·nm: the DNA double strands separate when a 9 pN·nm torque and >0.5 pN tension are applied against the double helix (Strick et al. 1999) and the ATP synthase $F_1$ motor generates a torque ~40 pN·nm MT.

2.3. Magnetic tweezers (MT)

AFM is composed of a microscopic cantilever with a stylus attached to one end, which reaches the surface of a biological sample (fig. 3a). The stylus interacts with the sample through the combination of Van der Waals attraction and Coulomb repulsion (fig. 3 b to e), resulting in deflection of the cantilever (Binnig et al. 1986). A laser beam is directed to the cantilever, reflects off the top surface and is imaged on to a split-photodiode that reads out the beam deflection. This way Angstrom resolution is readily achievable. AFM has two distinct applications: imaging and force measurement. In imaging mode, the cantilever performs a 2D raster scan over the sample surface and reads out the height profile. The high resolution stems from the fact that the imaging is not light based but rather mechanical contact mediated so far-field diffraction no longer applies. AFM does have the limit of low temporal resolution as the scanning tip has to physically reach each 'pixel' on the sample surface. In force measurement mode, the AFM tip no longer scans the sample surface but only moves vertically via a piezo electric actuator. The stylus is tethered to the biological molecule at one end and the chamber floor attaches to the other end. The cantilever can then be pulled to reveal a force-displacement curve, which reveals the conformational change of the biological sample.

3. Application of force spectroscopy tools in studying infection

In studying infections, force spectroscopy has been applied to a wide range of biological subjects: pathogenic proteins, nucleic acids and pathogenic/immune cell behaviours. Below we summarise a number of selected case studies to illustrate how the design of force spectroscopy tools allow them to achieve high resolution measurements of biological samples in physiological settings and how the results help progress our understanding in infection.

3.1. Resolving pathogenic protein actions

Nucleic acid proteins are vital to the health and functioning of all organisms. For instance, nucleic acids activities are facilitated with a large number of proteins or protein complexes. During DNA replication, topoisomerases (Champoux 2001) and gyrases (Gore et al. 2006) introduce both positive and negative twists into DNA to relieve extra turns built up from helicase unwinding and DNA local rotational constrains. Helicases (Patel and Donmez 2006) unwind and separate DNA double strands so DNA polymerases have access to the single strands for adding new nucleotides to form daughter double strands. During transcription, RNA polymerases create complementary RNA strands from single DNA strands (Murakami and Darst 2003). The list goes on. All of these activities and the mechanical properties of nucleic acids themselves are subjects of study for molecular tools.

RNA polymerases (RNAP) in bacteria are essential for nucleic acid transcription and can be inhibited resulting in bacterial death. In nature, bacteriophages have targeted the sigma factor in RNAP to modulate RNAP behaviours for the phages' benefit: *Xanthomonas oryzae* bacteriophage Xp10 encodes transcription regulator P7, an anti-sigma factor that displaces sigma factor during engagement of the RNAP with the promoter DNA. Sigma factor is an RNAP subunit that allows promoters to locate starting sites on DNA for transcription and phages modulate sigma so RNAPs transcribe their DNA rather than bacterial DNA (Liu et al. 2014). Understanding the mechanisms of RNAP actions inspires the development of antibacterial drugs. With optical tweezers, *Escherichia coli* RNA polymerase has been measured to move along DNA in discreet steps of $3.7 \pm 0.6$ Å (fig. 4 c and d) (Abbondanzieri et al. 2005), about the size of one DNA base pair, so they are shown to move one bp at a time. The sub-nm measurement resolution requires supreme beam stability from the optical tweezers used to make the measurement. The DNA being transcribed is tethered to a bead on one end. The RNAP is tethered to another bead while it moves along the DNA. The two beads are trapped separately with two optical traps – the 'dumbbell' configuration (fig. 4 a), which removed the need of tethering one end of the DNA to the reaction chamber (an alternative set up, fig. 4b), decoupling the system from the microscope stage thus eliminating stage drift. Also as the two traps arise from the same laser source by splitting the beam into two beams of orthogonal polarisation so laser pointing fluctuations do not result in relative shift in the two traps. Another measure to minimise the noise level is by sheltering the instrumentations in a helium-filled box to reduce interaction between air density fluctuations with the laser beam. This reduces low frequency noises to 1/100. The single molecule OT provides the unique capability to controllably apply a force to the RNAP to either assist its advancement along the DNA or to hinder it. This allowed the authors to plot RNAP movement velocity as a function of applied force, at a few different NTP concentrations. These relationships let the researchers deduce the kinetic model for RNA translocation. Existing competing models include Brownian ratchet mechanism (Schafer et al. 1991) in which the RNAP is free to diffuse both forward and backward along the DNA but incoming NTP acts as a ratchet to favour the forward motion and power stroke mechanism (Jiang and Sheetz 1994) in which the incorporation of NTP releases pyrophosphate that provides energy to RNAP translocation. The two models predict distinguished velocity-force plot and the assay verifies that the measured plot matches the Brownian ratchet mechanism and not the power stroke mechanism.

Hepatitis C virus (HCV) helicase as a drug target has captured much attention (Steimer and Klostermeier 2012) as deactivation of this enzyme disables all the nuclei acid processes that require unwinding and double-strand splitting thus halting viral replication. With the energy from ATP hydrolysis, HCV helicase moves along RNA or DNA to remove proteins bound to the nucleic acid and to unzip any complementary strands. A detailed understanding of the mechanism of its operation will facilitate the development of effective HCV helicase inhibitors. Despite much research attention, the mechanism of HCV helicase translocation is not fully understood due to the lack of spatial and temporal resolution in experimental measurements until single molecular measurements. X-ray crystallography reveals detailed HCV helicase structural information but it does not provide any information on the mechanistic cycles in which HCV helicase movement and ATP hydrolysis are coupled. Using optical tweezers, the series of actions have been resolved (Dumont 2006). An RNA hairpin was attached to two beads on both ends each via a DNA linkage (fig. 5a). One bead is tethered to an optical trap to allow measurement of force and the other bead to a micropipette for mechanical sub-nanometre position control. With active position feedback, the instrument can adjust the separation between the two beads to maintain either constant force levels or changing force levels. Thus, the separation of duplex RNA can be monitored at various pulling forces (fig. 5b), with the presence or absence of HCV helicase and different ATP concentrations. This assay allowed the measurement of movement in steps of multiples of $11 \pm 3$ bp and unwinding steps of multiples of $3.6 \pm 1.3$ bp (fig. 5c). The authors could conclude from their measurements the action model in which the translocator site on the helicase contacts the DNA 11 bp ahead of the helix opener site. The helicase advances in an inchworm fashion while the trailing helix opener separates the double strands.

As part of virus replication, DNA packaging motors push DNA into pre-assembled virus capsid against the tremendous internal forces due to the near-crystalline DNA density. Using optical tweezers, Smith *et al.* could measure the speed, pauses, slips and forces of packaging and establish the speed dependence on loadings (Liphardt et al. 2001). In their set-up, a bacteriophage $\varphi 29$ portal motor packages double-stranded DNA into the phage capsid (fig 6a). The capsid is tethered onto a bead whose position is fixed with a micropipette. The loose end of the DNA attaches to another bead trapped in the OT. The measurements are carried out in two modes: (1) constant force feedback mode where the tension in the DNA is kept at 5 pN by adjusting the bead separation (fig 6 b and c) and (2) zero feedback mode where the bead separation stays stationary and tension in DNA increases as the motor protein packages DNA (fig 6 d). In constant feedback mode, the packaging rate initially stayed constant but decreased quickly as the capsid filled up, indicating that the packaging rate is sensitive to internal pressure. 5 µm long DNA strand packaging have been observed, indicating high processivity. In zero feedback mode, the force builds up to about 55 pN before the motor stalls – the maximum motor force output. This also indicates the level of internal force due to DNA packaging. Small parts of packaged DNA also occasionally slips out of the capsid, shown as abrupt drops in force that lasts about 0.014 s – this also provides evidence that there is little resistance to DNA movement and thus little energy dissipation. Finally for DNA ejection, the high internal force at dense packing supports passive ejection for early stages but motor assisted ejection for latter stages.

3.2. Resolving inhibited protein actions

OT has also been applied to the characterisation of the actions of inhibited RNAP. Antibacterial peptide Microcin J25 (MccJ25) is produced by strains of *Escherichia coli* that binds to the secondary channel of RNAP such that nucleotide substrates cannot enter the enzyme active site. Adelman *et al*. first used biochemical assays and phosphor imaging (an ensemble biochemical method that quantifies and localises radioactively labelled nucleic acids in sample preparations such as gels and blots) to study the effect of MccJ25 on RNAP transcription and concluded that MccJ25 significantly inhibits transcription elongation even at saturating NTP concentrations and increases the appearance of transcriptional pauses. However, their assays could not quantify the reduction of elongation rate, nor could they distinguish whether MccJ25 caused additional pauses or extended the duration of existing pauses. Also, the complexity of the abortive initiation process made it difficult to extract mechanical details of the inhibition. The authors therefore resorted to optical tweezers that probe a single elongation complex with MccJ25 binding at a time to reveal the aforementioned parameters. In their OT set up, one end of a DNA substrate is tethered to a trapped micro-bead via streptavidin-bio link. The transcription site on the DNA binds to an RNAP which in turn is tethered to the chamber surface via (anti-HA)-(HA-epitope) link (fig. 7a). The OT and the DNA section between the bead and the RNAP is pulled taught such that any advancement of RNAP along the DNA manifests as the change in DNA length between the bead and RNAP. With feedback control, the tension in DNA is kept constant. Thus RNAP movement can be measured with interferometry bead displacement tracking in real time. Apart from consistent conclusions with biochemical studies over the decrease of RNAP processivity with increasing concentration of MccJ25, the optical tweezers assay also reveals through the elongation-time plot (fig. 7b) of individual transcription events that MccJ25 inhibition leads to increases in both frequency and duration of transcriptional pauses but the rate of 12.5 nt/s was not affected (fig. 7c). This suggests that MccJ25 stops rather than slows transcription: when it blocks secondary channels, it blocks completely, since partially blocked channels will manifest as slower transcription rates. Also since the timing of pauses are determined as well, the authors could find out that the stopping events are randomly distributed (Adelman et al. 2004). These findings are not accessible via ensemble detection methods.

The inhibition of DNA repairing protein RecA in *Mycobacterium tuberculosis* has been studied with magnetic tweezers. When double stranded DNA is broken, one of the repair mechanisms, homologous recombination, pairs one of the damaged strands with the corresponding section on the complementary strand of a similar or identical DNA ('the recipient DNA'), such as a sister chromatin, in order for the synthesis of a new strand to replace the damaged strand. In this process, RecA polymerises to form filament that bind to the single strand to assist the process. The regulatory protein RecX inhibits the polymerisation and promotes the depolymerisation of RecA filaments. Meanwhile, mechanical force arises from the stretching of the RecA-ssDNA complex as the complex translocates to reach the recipient DNA. This force reduces the inhibition of RecA filament due to RecX. Using MT, Le *et al*. characterised the responses of the inhibitory effects to different force levels and various concentrations of RecX (Le et al. 2014). A DNA is tethered vertically between a magnetic bead and the coverslip surface. Two magnets with opposite poles are held above the magnetic bead to provide the upward pulling force (fig. 2a and 8a). With video imaging and defocusing diffraction pattern analysis, the vertical position of the bead can be determined with nanometre resolution. Force-extension curves are plotted for naked ssDNA and full-length RecA-bound ssDNA (fig. 8b). The latter shows much higher rigidity: naked DNA is initially coiled and extends rapidly as tension increases but RecA bound DNA is extended initially and does not stretch much further as force increases. This establishes a method to determine the extent of RecA filament polymerisation and depolymerisation: it varies the vertical bead position.  At low forces, RecX inhibits RecA filament formation and encourages depolymerisation (fig. 8c and d) but this effect is reduced at higher applied tension in the filament to the extent that partially disassembled filament can re-grow (fig. 8e). This suggests a mechanosensitive regulation mechanism: during homologous recombination, the tension in DNA-RecA complex increases. To keep both ends of the broken DNA together, the DNA-RecA complex stabilises as a response to the tension.

3.3. Investigating immune cells

Ashkin et al. showed OT capability of trapping bacteria, viruses (Ashkin and Dziedzic 1987) and human cells (Ashkin and Dziedzic 1987) as early as 1987. In contrast, due to the lack of ferro-/para-magnetic properties, biological samples cannot be directly manipulated by MT, but a magnetic microparticle can be inserted into a cell and the magnetic translation of this particle perturbs the structure and dynamics of the cell to reveal cell properties – magnetic microrheology. Alternatively, a trapped magnetic particle can be brought into contact of cells to probe adhesion properties.

Macrophages are immune cells that engulf and/or destroy any objects in the body that are not recognised as healthy body cells for not having the right type of surface signalling proteins, potentially because they are invaders, pathogens in particular. The understanding of these cells is crucial to understanding the immune system. The viscoelasticity of the macrophage cytoplasm has direct implications on cell motion and deformation, intracellular transportation and phagocytosis, all of which have direct impacts on the macrophages' immune response. Further, viscoelastic parameters allow quantitative characterization of the structure of the cytoplasm. Cytoplasmic viscoelastic moduli are determined from viscoelastic creep curve: a constant local stress applied on the material for a period of time deforms the sample (the deformation known as strain) and the stress-strain vs time is plotted. The curve consists of three qualitatively distinctive parts: a fast elastic response followed by a slower response to the applied stress and finally relaxation when the stress is removed. From the curve, viscosity, relaxation time and shear constant can be calculated. Magnetic tweezers are used to apply the stress: micron-sized magnetic beads are phagocytized by macrophages. The magnetic force to probe the cytoplasm needs to reach nanonewtons, higher than usual MT designs are capable of. But this can be done with electromagnetic tweezers by having many coil turns, large electric current and high magnetic core permeability whilst bringing the core close to the sample. Bausch et al. devised a field generation unit composing of a single coil with 1200 turns wound round a soft iron core that supports up to 4A current (Bausch et al. 1998). The step force is in one direction only so one coils suffices. This allows the tip of the core piece to reach as close as 10 μm to the sample. With CCD camera video microscopy and single particle tracking algorithm, the spatial resolution reached 10 nm and temporal resolution 0.04 s (Bausch et al. 1999). The authors measured a shear modulus ranging from 20-735 Pa, viscosity 210 ± 143 Pa·s. Also non-magnetic beads are tracked to characterise the displacement field and the authors concluded that the cytoplasm is composed of densely packed filaments with soft gaps. They also could measure local active forces within the cytoplasm potentially due to motor proteins or local flow of cytoplasm. Prior methods to measure forces have resorted to averaging during long periods of time and have been hampered by the high level of heterogeneity in the cytoplasm of macrophages.

3.4. Characterising virus attachment to host cells

*Plasmodium*, commonly known as malaria parasites, is a genus of unicellular eukaryotic organisms that have more than 100 species, many of which can infect humans and animals to cause malaria. The parasites found in mosquito saliva are at a reproductive stage when they are called *sporozoites* and are capable of infecting new hosts. Following a mosquito bite, the sporozoites are deposited under the skin of the recipient and they migrate to enter the host's circulation system and finally the liver where the parasites move on to the next stage of life cycle. Cellular mobility is essential to plasmodium reproduction and infection of their hosts. Sporozoites have the shape of a crescent (fig. 9). The cell surface features adhesion-mediating proteins for sites that adhere to cellular substrates in physiological environments but also to carbon and gold surfaces in experimental settings. When sporozoites navigate along a surface, new adhesion sites form in the leading edge and existing sites disassemble at the trailing edge. The complex interplay between surface proteins and signalling events involved in this gliding motility is made accessible via optical tweezers manipulation of the parasites. Hegge *et al*. devised an OT assay capable of trapping and manipulating a part of the crescent parasite and characterising the forces that the laser trap applies to the cell (Hegge et al. 2012). To avoid blocking adhesion sites on the surface of the sporozoites, the sporozoites are trapped directly rather than via an attached bead. The major challenge lies in the non-spherical shape of the cells as well as the heat damage to the cells. The long axis of the sporozoites exceeds the extent of the trap focus so only a part of the cell is trapped and manipulated. The cells are soft and thus bend under the force so trap movements of a few microns do not necessarily move the cells. This means standard thermal noise characterisation of the trap stiffness does not apply. The authors use flow pressure to push the cell out of the trap during which the escape velocity and thus trap stiffness can be determined. The attaching and locomotion behaviour of parasites that are genetically modified to lack specific proteins are manipulated to uncover the distinct roles that each protein plays as well as the interaction between species of proteins. For example, it was established that both the surface proteins TRAP and S6 are required for correct initial adhesion, that S6 determines the strength of adhesion for resistance against externally applied axial force (fig. 9 d to g) and that TRAP determines resistance against rotation (fig. 9c).

Mammen *et al.* devised the 'OPTCOL' (optically controlled collision) assay to measure the probability of adhesion of viruses to host cells upon collision. OPTCOL is a pair of independently controlled traps (the 'dumbbell' configuration) in which one trap manipulates the host cell and the other manipulates a 5 μm bead whose surface is coated with a virus (Mammen 1996). The traps first move towards each other to bring the virus into contact with the host cell and then the traps move away so adherence/separation between the two bodies can be determined by observing if the two bodies attach or separate. Statistics gathered are used to calculate the probability of attachment at various inhibitor concentrations. Previous ensemble methods suspend virus and host cells at certain concentrations in the solution. Due to size differences, host cells fall to the bottom of reaction chamber by gravity while viruses do not. This creates collision events between the two bodies. Attachments between multiple hosts and viruses eventually form a macroscopic aggregate. The limitation of this method is that inhibitor concentration has a minimum measurable value below which different types of inhibitors appear to be equally effective. The significance of OT compared to ensemble methods, apart from measuring inhibitor effectiveness at extremely low concentrations, includes the ability to vary collision velocity which mimics different situations inside human bodies and which has potent influence over attachment statistics, the capability to extend agents of interest to a wide range of mesoscopic objects and provide detailed adhesion distributions over a collection of individual collisions.

4. Conclusion

Introduced in the 1980s, force spectroscopy is still in its infancy. In recent years, advancements in components such as lasers and CMOS camera that make up force spectroscopy instruments have allowed the increase of resolution in force spectroscopy instruments. In addition, combinations of spectroscopy tools with each other allow unique manipulation and measurement capabilities so a wider range of biological systems can be probed. For example, magneto-optical tweezers (Crut 2007; Zhou et al. 2015) are a chimera of OT and MT that can manipulate particles in all translational and rotational degrees of freedom without the technical challenges in implementing rotation-enabled OT (La Porta and Wang 2004) as MT easily applies twists or in implementing high resolution MT (Dulin et al. 2015) as OT intrinsically has high resolution. Combination of force spectroscopy such as super-resolution microscopy adds direct optical visualisation to molecular activities (Miller et al. 2015). Another area that will likely see significant development is device multiplicity in the sense that multiple copies of biological samples are simultaneously experimented on high throughput devices. Due to heterogeneity among biological samples, force spectroscopy investigations need to be statistical in nature. Holographic OT (Padgett and Di Leonardo 2011) and DNA curtains (Greene et al. 2010) are some examples of configuration and device designs to allow multiplicity. Many biological systems relevant to infection remain to be investigated and, with the knowledge already obtained with ensemble methods, experimental results from force spectroscopy will prove to be useful.

Abbondanzieri EA, Greenleaf WJ, Shaevitz JW, Landick R, Block SM (2005) Direct observation of base-pair stepping by RNA polymerase. Nature 438:460-465

Adelman K, Yuzenkova J, La Porta A, Zenkin N, Lee J, Lis JT, Borukhov S, Wang MD, Severinov K (2004) Molecular mechanism of transcription inhibition by peptide antibiotic Microcin J25. Mol Cell 14 (6):753-762. doi:10.1016/j.molcel.2004.05.017

Ashkin A (1992) Forces of a single-beam gradient laser trap on a dielectric sphere in the ray optics regime. Biophysical journal 61 (2):569-582. doi:10.1016/S0006-3495(92)81860-X

Ashkin A, Dziedzic JM (1987) Optical trapping and manipulation of viruses and bacteria. Science 235 (4795):1517-1520

Bausch AR, Moller W, Sackmann E (1999) Measurement of local viscoelasticity and forces in living cells by magnetic tweezers. Biophys J 76:573-579

Bausch AR, Ziemann F, Boulbitch AA, Jacobson K, Sackmann E (1998) Local measurements of viscoelastic parameters of adherent cell surfaces by magnetic bead microrheometry. Biophys J 75:2038-2049

Binnig G, Quate CF, Gerber C (1986) Atomic force microscope. Phys Rev Lett 56 (9):930-933. doi:10.1103/PhysRevLett.56.930

Block SM, Blair DF, Berg HC (1989) Compliance of bacterial flagella measured with optical tweezers. Nature 338 (6215):514-518. doi:10.1038/338514a0

Block SM, Goldstein LS, Schnapp BJ (1990) Bead movement by single kinesin molecules studied with optical tweezers. Nature 348:348-352

Bullard B, Garcia T, Benes V, Leake MC, Linke WA, Oberhauser AF (2006) The molecular elasticity of the insect flight muscle proteins projectin and kettin. Proceedings of the National Academy of Sciences of the United States of America 103 (12):4451-4456

Champoux JJ (2001) DNA topoisomerases: structure, function, and mechanism. Annual review of biochemistry 70:369-413. doi:10.1146/annurev.biochem.70.1.369

Charvin G, Strick TR, Bensimon D, Croquette V (2005) Tracking topoisomerase activity at the single-molecule level. Annu Rev Biophys Biomol Struct 34:201-219

Crut AK, D. A.; Seidel, R.; Wiggins, C. H.; Dekker, N. H. (2007) Fast dynamics of supercoiled DNA revealed by single-molecule experiments. Proceedings of the National Academy of Sciences of the United States of America 104 (29):11957-11962. doi:10.1073/pnas.0700333104

De Vlaminck ID, C. (2012) Recent advances in magnetic tweezers. Annual review of biophysics 41:453-472. doi:10.1146/annurev-biophys-122311-100544

Dulin D, Cui TJ, Cnossen J, Docter MW, Lipfert J, Dekker NH (2015) High Spatiotemporal-Resolution Magnetic Tweezers: Calibration and Applications for DNA Dynamics. Biophysical journal 109 (10):2113-2125. doi:10.1016/j.bpj.2015.10.018

Dumont S (2006) RNA translocation and unwinding mechanism of HCV NS3 helicase and its coordination by ATP. Nature 439:105-108

Fili N, Mashanov GI, Toseland CP, Batters C, Wallace MI, Yeeles JTP, Dillingham MS, Webb MR, Molloy JE (2010) Visualizing helicases unwinding DNA at the single molecule level. Nucleic acids research 38 (13):4448-4457. doi:10.1093/nar/gkq173

Finer JT, Simmons RM, Spudich JA (1994) Single myosin molecule mechanics: piconewton forces and nanometre steps. Nature 368 (6467):113-119. doi:10.1038/368113a0

Gennerich A, Carter AP, Reck-Peterson SL, Vale RD (2007) Force-induced bidirectional stepping of cytoplasmic dynein. Cell 131 (5):952-965. doi:10.1016/j.cell.2007.10.016

Gore J, Bryant Z, Stone MD, Nöllmann M, Cozzarelli NR, Bustamante C (2006) Mechanochemical analysis of DNA gyrase using rotor bead tracking. Nature 439 (7072):100-104

Gosse C, Croquette V (2002) Magnetic tweezers: micromanipulation and force measurement at the molecular level. Biophys J 82:3314-3329

Greene EC, Wind S, Fazio T, Gorman J, Visnapuu ML (2010) DNA curtains for high-throughput single-molecule optical imaging. Methods in enzymology 472:293-315. doi:10.1016/S0076-6879(10)72006-1

Hegge S, Uhrig K, Streichfuss M, Kynast-Wolf G, Matuschewski K, Spatz JP, Frischknecht F (2012) Direct manipulation of malaria parasites with optical tweezers reveals distinct functions of Plasmodium surface proteins. ACS Nano 6 (6):4648-4662. doi:10.1021/nn203616u

Jiang MY, Sheetz MP (1994) Mechanics of myosin motor: force and step size. Bioessays 16 (8):531-532. doi:10.1002/bies.950160803

Kellermayer MS, Smith SB, Granzier HL, Bustamante C (1997) Folding-unfolding transitions in single titin molecules characterized with laser tweezers. Science 276 (5315):1112-1116

King GA, Gross P, Bockelmann U, Modesti M, Wuite GJL, Peterman EJG (2013) Revealing the competition between peeled ssDNA, melting bubbles, and S-DNA during DNA overstretching using fluorescence microscopy. Proceedings of the National Academy of Sciences of the United States of America 110 (10):3859-3864. doi:10.1073/pnas.1213676110

La Porta A, Wang MD (2004) Optical torque wrench: angular trapping, rotation, and torque detection of quartz microparticles. Phys Rev Lett 92 (19):190801

Le S, Chen H, Zhang X, Chen J, Patil KN, Muniyappa K, Yan J (2014) Mechanical force antagonizes the inhibitory effects of RecX on RecA filament formation in Mycobacterium tuberculosis. Nucleic acids research 42 (19):11992-11999. doi:10.1093/nar/gku899

Leake MC (2013) The physics of life: one molecule at a time. Philosophical Transactions of the Royal Society B: Biological Sciences 368 (1611):20120248

Leake MC, Grützner A, Krüger M, Linke WA (2006) Mechanical properties of cardiac titin's N2B-region by single-molecule atomic force spectroscopy. Journal of structural biology 155 (2):263-272

Leake MC, Wilson D, Bullard B, Simmons RM (2003) The elasticity of single kettin molecules using a two-bead laser-tweezers assay. Febs Lett 535 (1-3):55-60. doi:Doi 10.1016/S0014-5793(02)03857-7

Leake MC, Wilson D, Gautel M, Simmons RM (2004) The elasticity of single titin molecules using a two-bead optical tweezers assay. Biophysical journal 87 (2):1112-1135. doi:DOI 10.1529/biophysj.103.033571

Lidstrom ME, Konopka MC (2010) The role of physiological heterogeneity in microbial population behavior. Nat Chem Biol 6 (10):705-712. doi:10.1038/nchembio.436

Linke WA, Leake MC (2004) Multiple sources of passive stress relaxation in muscle fibres. Physics in medicine and biology 49 (16):3613

Liphardt J, Onoa B, Smith SB, Tinoco I, Bustamante C (2001) Reversible unfolding of single RNA molecules by mechanical force. Science 292:733-737

Liu B, Shadrin A, Sheppard C, Mekler V, Xu Y, Severinov K, Matthews S, Wigneshwerarj S (2014) A bacteriophage transcription regulator inhibits bacterial transcription initiation by sigma-factor displacement. Nucleic acids research 42 (7):4294-4305. doi:10.1093/nar/gku080

Mammen M (1996) Optically controlled collisions of biological objects to evaluate potent polyvalent inhibitors of virus-cell adhesion. Chem Biol 3:757-763

Miller H, Zhou Z, Wollman AJ, Leake MC (2015) Superresolution imaging of single DNA molecules using stochastic photoblinking of minor groove and intercalating dyes. Methods. doi:10.1016/j.ymeth.2015.01.010

Moffitt JR, Chemla YR, Smith SB, Bustamante C (2008) Recent advances in optical tweezers. Annual review of biochemistry 77:205-228. doi:10.1146/annurev.biochem.77.043007.090225

Mosconi FA, J. F.; Croquette, V. (2011) Soft magnetic tweezers: a proof of principle. The Review of scientific instruments 82 (3):034302. doi:10.1063/1.3531959

Murakami KS, Darst SA (2003) Bacterial RNA polymerases: the wholo story. Current opinion in structural biology 13 (1):31-39

Neuman KC, Nagy A (2008) Single-molecule force spectroscopy: optical tweezers, magnetic tweezers and atomic force microscopy. Nature methods 5 (6):491-505. doi:10.1038/nmeth.1218

Padgett M, Di Leonardo R (2011) Holographic optical tweezers and their relevance to lab on chip devices. Lab Chip 11 (7):1196-1205. doi:10.1039/c0lc00526f

Patel SS, Donmez I (2006) Mechanisms of helicases. J Biol Chem 281 (27):18265-18268. doi:10.1074/jbc.R600008200

Schafer DA, Gelles J, Sheetz MP, Landick R (1991) Transcription by single molecules of RNA polymerase observed by light microscopy. Nature 352 (6334):444-448. doi:10.1038/352444a0

Shevkoplyas SS, Siegel AC, Westervelt RM, Prentiss MG, Whitesides GM (2007) The force acting on a superparamagnetic bead due to an applied magnetic field. Lab Chip 7 (10):1294-1302. doi:10.1039/b705045c

Smith SB, Cui Y, Bustamante C (1996) Overstretching B-DNA: the elastic response of individual double-stranded and single-stranded DNA molecules. Science 271 (5250):795-799

Smith SB, Cui YJ, Bustamante C (2003) Optical-trap force transducer that operates by direct measurement of light momentum. Method Enzymol 361:134-162

Steimer L, Klostermeier D (2012) RNA helicases in infection and disease. Rna Biol 9 (6):751-771. doi:10.4161/rna.20090

Strick TR, Allemand JF, Bensimon D, Bensimon A, Croquette V (1996) The elasticity of a single supercoiled DNA molecule. Science 271:1835-1837

Strick TR, Allemand JF, Bensimon D, Croquette V (1998) Behavior of supercoiled DNA. Biophys J 74:2016-2028

Strick TR, Bensimon D, Croquette V (1999) Micro-mechanical measurement of the torsional modulus of DNA. Genetica 106 (1-2):57-62

Zhou Z, Miller H, Wollman A, Leake M (2015) Developing a New Biophysical Tool to Combine Magneto-Optical Tweezers with Super-Resolution Fluorescence Microscopy. Photonics 2 (3):758

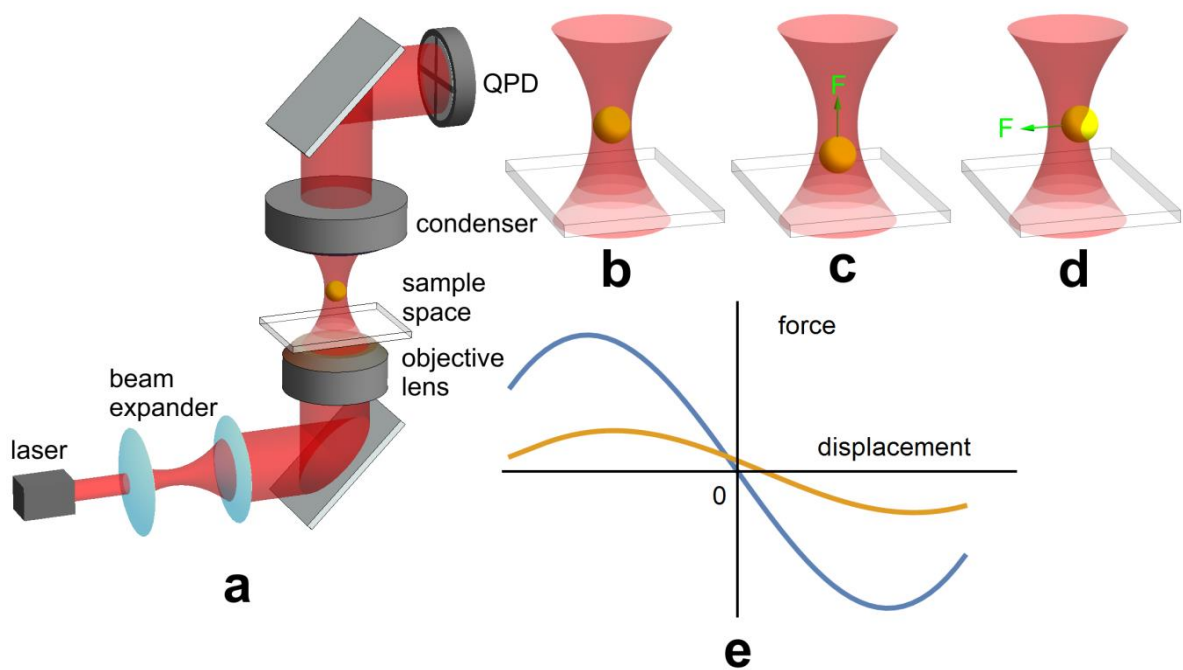

**Fig. 1**

**Fig. 1** The force transduction mechanism of optical tweezers (OT). **a** An overview of a basic OT. The laser emits a collimated infrared beam that is expanded with a telescope beam expander to overfill the objective lens. The objective lens focuses the laser beam to a diffraction-limited spot at the beam waist (the narrowest place of the beam). The trapped bead scatters the beam and creates an inference pattern at the back focal plane of the condenser, which is imaged via a positive lens (not shown) at a quadrant photo diode to determine the bead position with sub-nanometre resolution. **b** shows the bead in equilibrium position – along the beam axis and slightly past the point of highest intensity. Apart from trapping force, there is also a scattering force due to the absorption of some photons and thus the absorption of their forward moving momenta, so the bead is actually slightly

past the beam waist. In the main text we have ignored this effect for simplicity. When the bead is displaced downward as shown in **c** and sideways as shown in **d**, there appears a restoring force pushing the bead towards the equilibrium position, represented by the green arrow and the letter F. **e** plots force-displacement curves for axial (orange) and radial (blue) bead displacements from the centre of the trap. The relationship is approximately linear close to the centre, allowing calculation of trapping force from measured displacement.

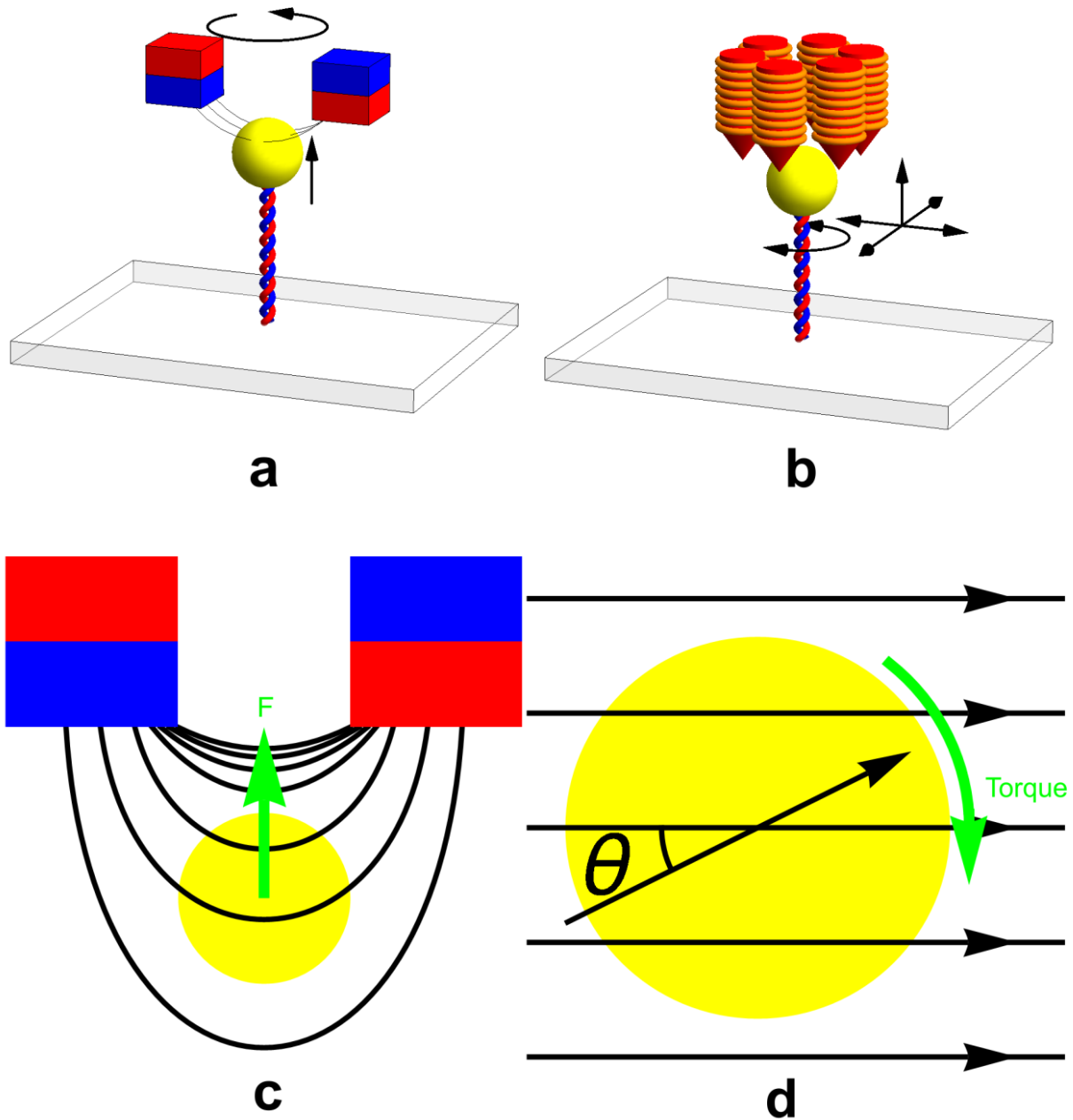

**Fig. 2**

**Fig. 2** The force and torque transduction mechanism of magnetic tweezers (MT). The B field is created either by permanent magnets (**a**) or electromagnets (**b**). **a** shows an experimental assay where a double-stranded DNA is tethered to the cover slip at the bottom and a bead at the top. The bead can be pulled and twisted by moving the two magnets: moving the magnets closer to the bead

increases the field gradient at the bead so increases the magnetic force; rotating the magnets will rotate the bead. **b** is the electromagnet equivalent that features six coils (Gosse and Croquette 2002) and has the added capability of applying horizontal forces compared to the configuration in **a**. A magnetic bead in this ***B*** field is subject to a magnetic force proportional to the ***B*** field gradient as shown in **c**. The force is indicated by the green arrow. Like OT in which a bead has the tendency of moving towards laser intensity maximum, the magnetic bead tends to move towards the ***B*** field maximum. **d** illustrates the mechanism that gives rise to magnetic torques: misalignment of the magnetisation of the magnetic bead from the background ***B*** field, shown as the angle $\vartheta$. The torque is indicated by the curved green arrow.

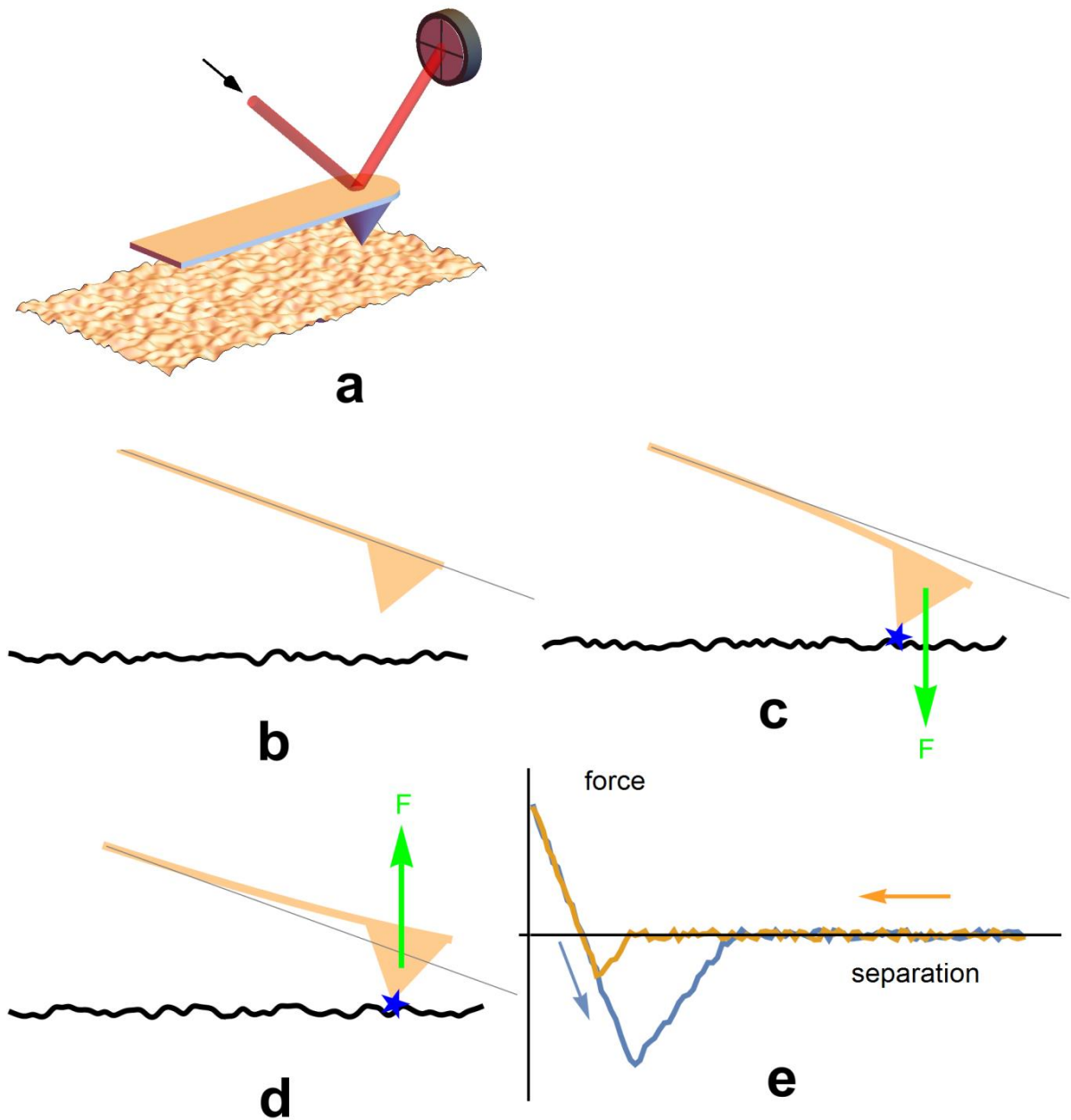

**Fig. 3**

**Fig. 3** Atomic force microscope (AFM) and its imaging mechanisms. **a** Schematics of AFM imaging a surface profile. The cantilever has a stylus attached to one end. The stylus is brought into physical contact with the sample surface. When the cantilever scans the surface, the local height changes mechanically move the stylus vertically thus deflecting the cantilever. A laser beam reflects off the upper face of the cantilever and a split-photodiode (SPD) catches the reflected light, which allows the measurement of the deflection. **b** - **d** show the force levels that the surface transduces to the cantilever and the bending of the cantilever. The green arrows represent forces. In **b** the separation between the cantilever and the substrate is large and there is no interaction. In **c** the stylus is close enough to the substrate so that a Van der Waals attraction bends the stylus downward. **d** As the cantilever further approaches the substrate, a Coulomb repulsion develops that overcomes the Van der Waals attraction and the cantilever feels a push upward. **e** shows an idealised representation of the force changes as the stylus approaches and leaves the substrate, as indicated by the orange and blue curves respectively. The dents in the curves are due to Van der Waals attraction.

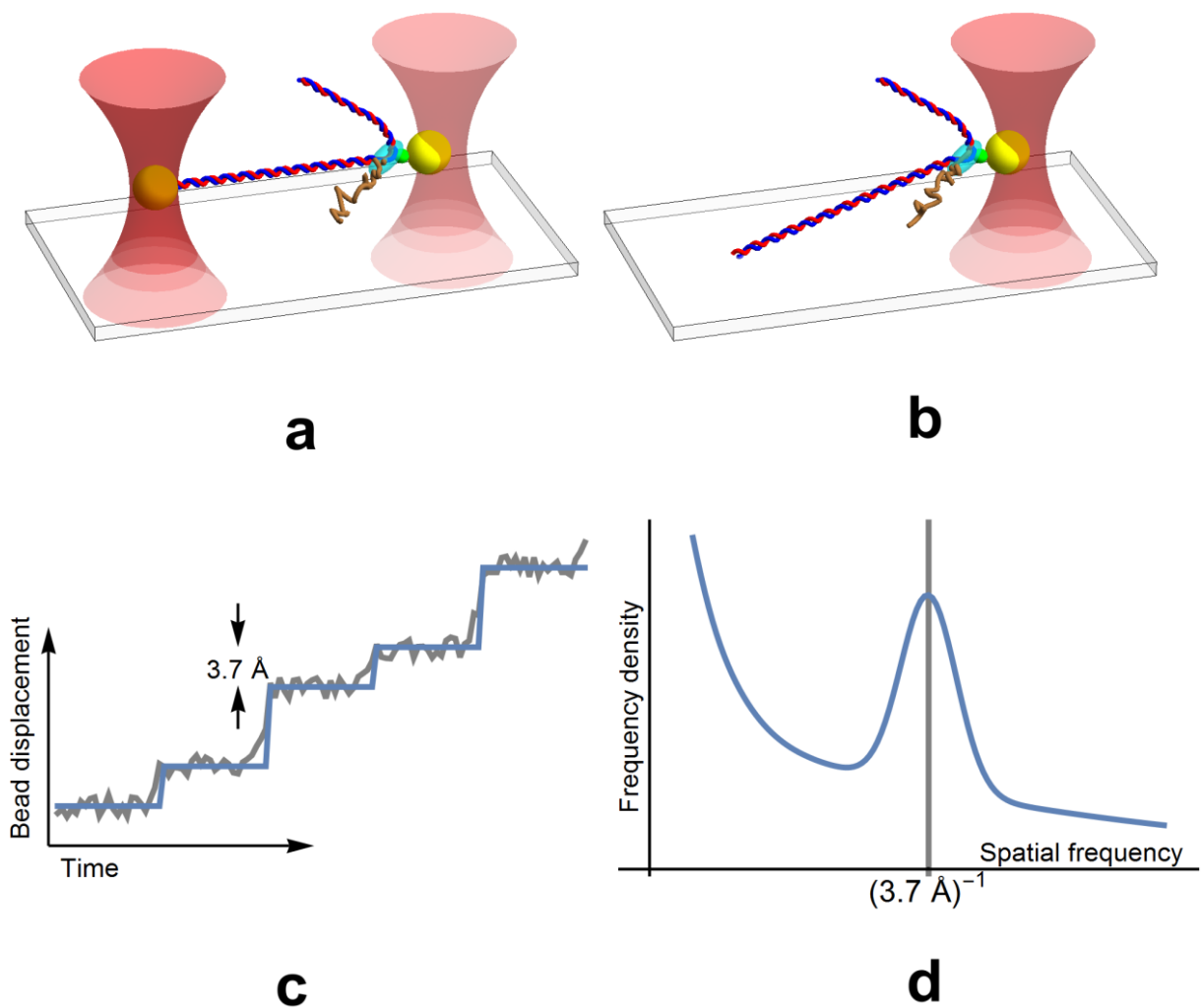

**Fig. 4**

**Fig. 4** Optical trap set-up to study RNAP activity and schematics of results. **a** Two laser traps are created at the sample space to trap two beads simultaneously. The advantage of this configuration over an alternative single trap design in which one end of the DNA is fixed to the surface of the coverslip as shown in **b** is the avoidance of any mechanical noise or drift from the coverslip from transducing to the biological sample itself thus better resolution can be achieved. **c** A schematic showing the displacement over time measurement that illustrates the identification of 3.7 Å step size. **d** A spatial frequency plot of the data in **c** identifying a peak at (3.7 Å)$^{-1}$.

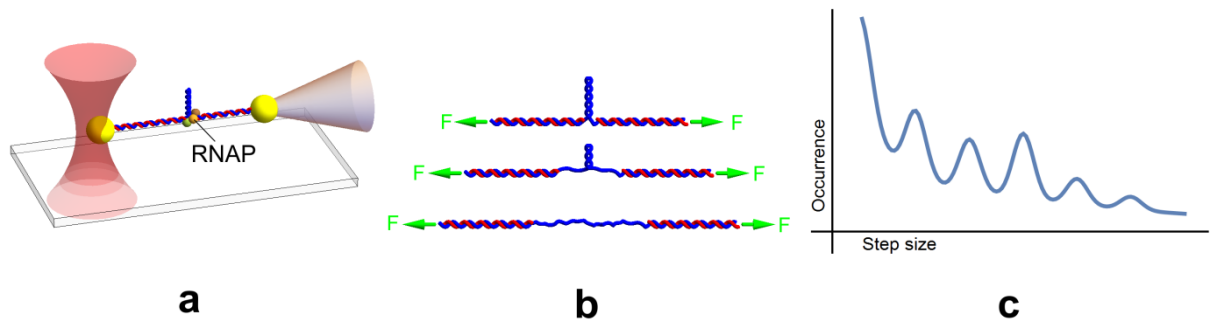

**Fig. 5**

**Fig. 5** OT assay to probe the mechanical cycles of HCV NS3 helicase. **a** the OT configuration featuring an RNA hairpin (blue) tethered on both ends to beads via ssDNA handles (red). One end is trapped in OT with active feedback allowing constant-force mode. The other end is fixed onto a pipette tip with sub-nanometre mechanical positioning. Remarkably, the trap is created with two counter propagating beams so the forward-going scattering forces due to the two beams are cancelled, resulting in stiffer trapping along the axial direction. This way lower numerical aperture objective lenses can be used and all the light leaving the trap can be measured and analysed to determine the rate of momentum change (Smith et al. 2003). **b** From top to bottom: the RNA hairpin denatures at catalysis of NS3 Helicase. The application of constant weak forces on both ends of the structure pulls the structure taut. Time traces of bead displacements such as the one shown in fig.4c can be determined from this assay, which allows the characterisation of step sizes shown in **c**. **c** discrete step sizes of various magnitudes are shown through the peaks in the step size versus occurrence plot.

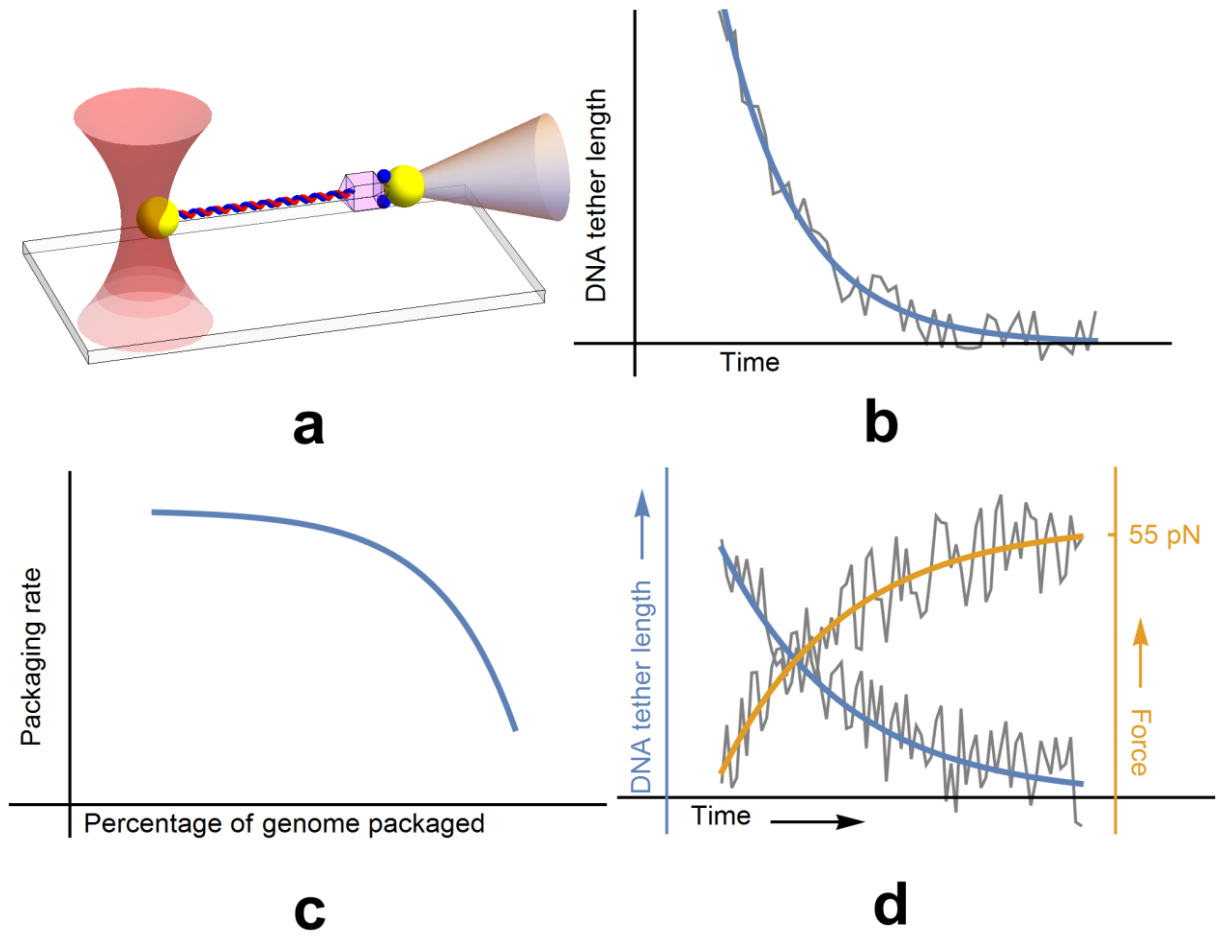

**Fig. 6**

**Fig. 6** OT assay to probe motor proteins that package DNA into virus capsid. **a** schematics of the assay: the capsid of the virus is fixed to a pipette tip while the other end of the DNA is trapped in the OT. **b** The constant-force mode. DNA outside the capsid is shortened over time as motor proteins package the genome into the capsid. **c** derives from data in **b** and shows the drop of packaging rate as the packaging progresses. **d** The assay operates in no-feedback mode so the separation between the trapping beam and the pipette tip stays stationary. The tension in DNA builds up as the motor protein pulls the DNA and packages it into the capsid. The protein halts at approximately 55 pN, suggesting the maximum pulling strength of this protein. This also hints on the internal pressure that can be built up via packaging. The authors argued that this pressure may be used for DNA ejection inside host bodies.

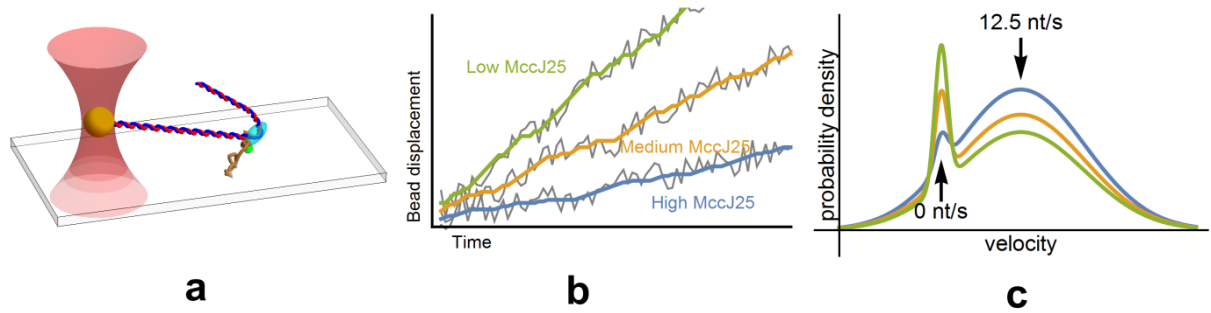

**Fig. 7**

**Fig. 7** OT assay to probe the inhibition of antibacterial peptide Microcin J25 (MccJ25) on the activity of bacterial RNA polymerase. **a** The assay set up differs from that in fig. 4a in that only one trap is used here. The RNAP end of the DNA is tethered to the camber surface, subject to the transmission of mechanical noise to the biological sample. This configuration is simpler to implement while providing enough resolution over the time scale of ~100 seconds to characterise the transcription rate. **b** shows time traces of RNAP advancements along DNA at three concentrations of the inhibitor MccJ25. Individual moving and pausing events can be identified as tilting and horizontal parts in the curve to allowing the authors to plot transcription rate distribution in **c**: it turns out that MccJ25 does not alter transcription rate but rather increases the frequency of pauses.

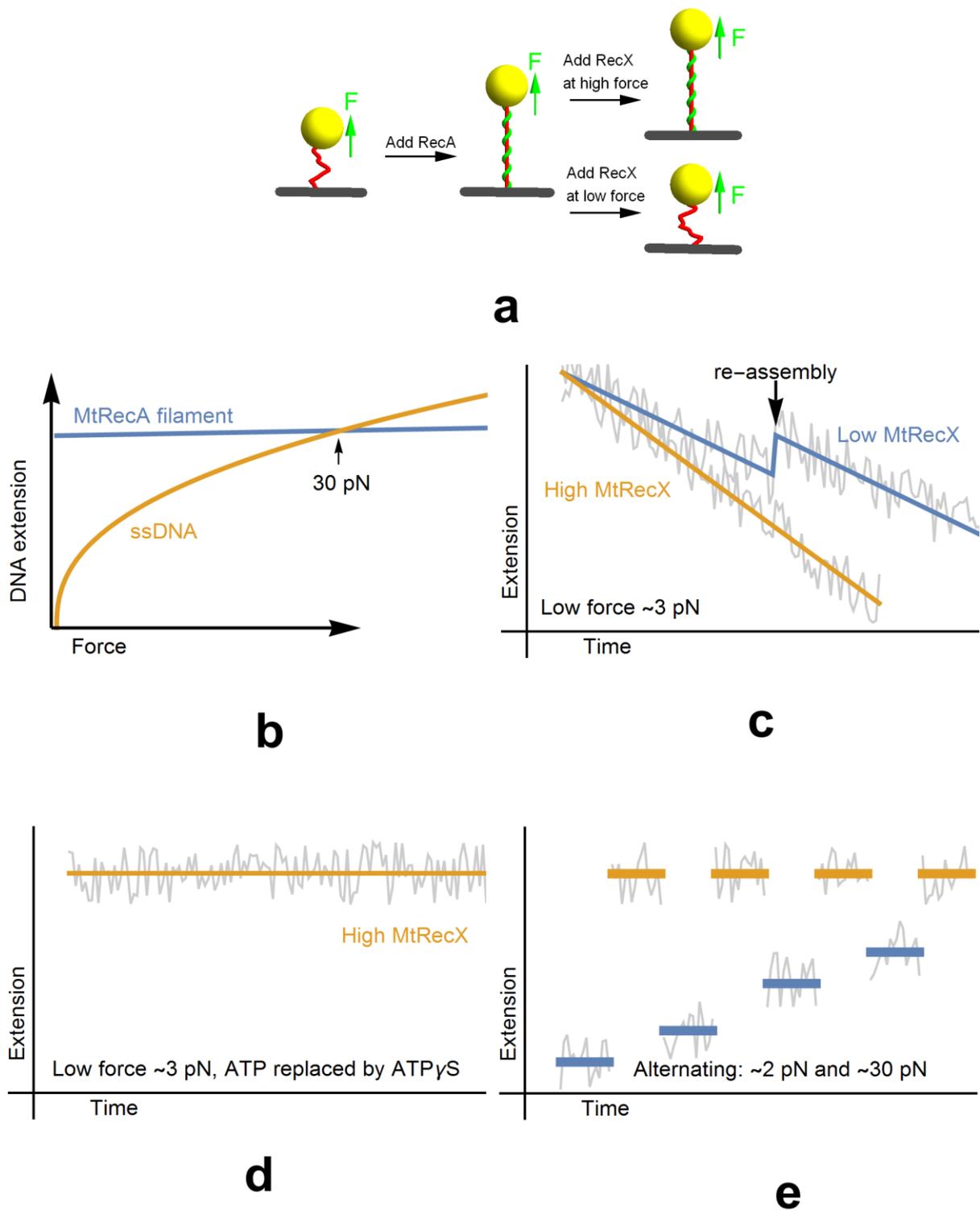

**Fig. 8**

**Fig. 8** Magnetic tweezers assay to probe RecX inhibition of RecA filament formation and decomposition. **a** A graphical representation of the main findings. The ssDNA (red curve) curls up at low force (left) but addition of RecA forms a rigid filament along the DNA, straightening the DNA thus pushing the bead away from the chamber surface (middle). Further addition of RecX depolymerises the RecA filament at low forces (right, bottom) but the inhibition effect is less prominent at high forces (right, top). **b** DNA extension vs force curves with and without RecA to show the difference in the rigidity of the two scenarios. This establishes a method to measure the amount of filament formation. **c** Time traces of extension showing RecA formation at low and high RecX respectively. **d** When ATP is replaced by non-hydrolysable homologue, ATPγS, no RecA filament decomposition is detected, pointing to ATP dependence of RecX inhibition. **e** Alternating low and high forces showing the roles that forces play on the effectiveness of RecX inhibition on RecA filament formation.

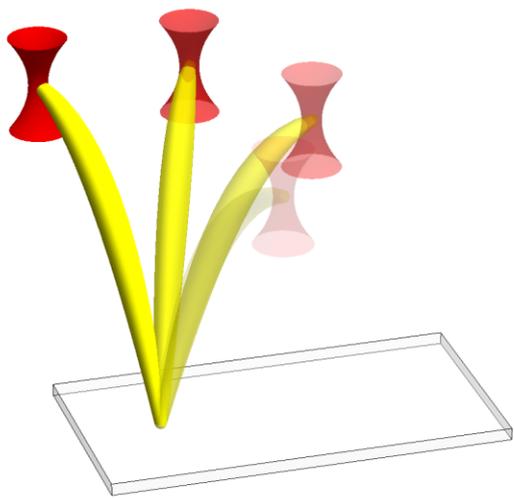

a

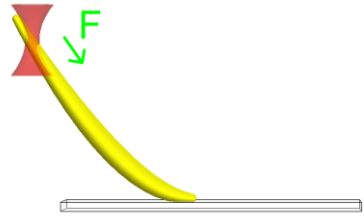

d

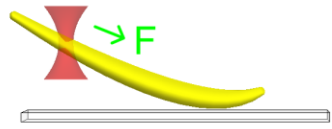

e

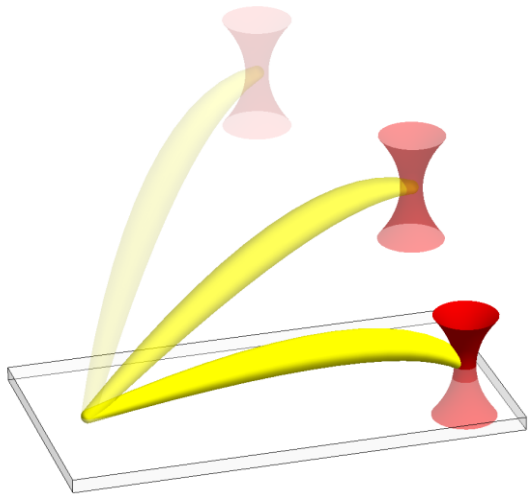

b

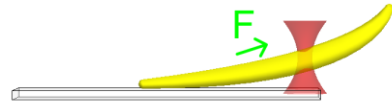

f

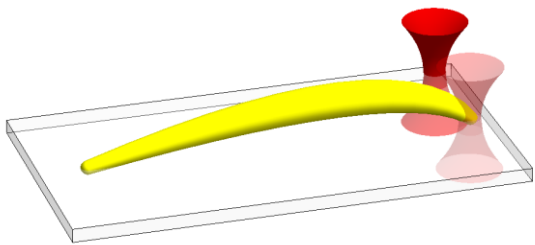

c

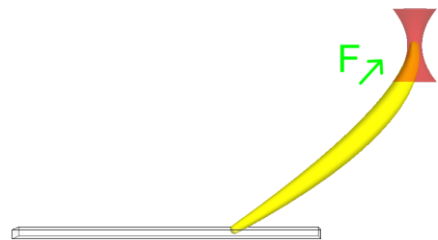

g

**Fig. 9**

Optical tweezers (OT) manipulate a malaria parasite. A few modes of rotating and moving the parasite are shown to illustrate the capability of OT in interacting with the parasite. **a** One end of the parasite adheres to the coverslip via its surface adhesion protein. The rest of the parasite is free from the glass so the parasite is not rotationally constrained. The other end of the parasite is trapped in the OT focus. The relative position of the trap with respect to the coverslip rotates around the adhesion point and parasite rotates with the trap. **b** the trap position relative to the coverslip is closed down to bring the free end of the parasite to the coverslip. **c** Once the free end touches the coverslip, adhesion sites may form and if the 100mW trap is moved away now, the parasite will not follow the trap, indicating strong adhesion. **d – g** When only one adhesion site is formed between the parasite and the coverslip and when the trap moves along the long axis, it is possible to push the parasite forward. The green arrows indicate the direction of pushing and pulling. During the movement, the adhesion site migrates along the parasite surface.